\documentclass[a4paper,11pt]{article}
\usepackage{pos}
\usepackage{natbib}
\usepackage{floatrow}

\title{Assessing the signatures imprinted by star-forming galaxies in the cosmic $\gamma$-ray background}
 \ShortTitle{Signatures of star-forming galaxies in the $\gamma$-ray background}

\author*[a,b]{Ellis R. Owen}
\author[c]{Khee-Gan Lee}
\author[a]{Albert K. H. Kong}

\affiliation[a]{Institute of Astronomy, National Tsing Hua University, Hsinchu, Taiwan (ROC)}
\affiliation[b]{Center for Informatics and Computation in Astronomy, National Tsing Hua University, \\Hsinchu, Taiwan (ROC)}
\affiliation[c]{Kavli Institute for the Physics and Mathematics of the Universe, University of Tokyo, Kashiwa, Chiba 277-8583, Japan}

\emailAdd{erowen@gapp.nthu.edu.tw}

\abstract{In recent years, $\gamma$-ray emission has been detected from star-forming galaxies (SFGs) in the local universe, including M82, NGC 253, Arp 220 and M33. The bulk of this emission is thought to be of hadronic origin, arising from the interactions of cosmic rays (CRs) with the interstellar medium of their host galaxy. Distant SFGs are presumably also bright in $\gamma$-rays. Although they would not be resolvable as point sources, distant unresolved SFG populations contribute $\gamma$-rays to the extra-galactic $\gamma$-ray background (EGB). Despite the wealth of high-quality all-sky EGB data 
collected over more than a decade of operation
with the \textit{Fermi}-LAT $\gamma$-ray space telescope, the exact contribution of SFGs to the EGB remains unsettled. In this study, we model the $\gamma$-ray emission from SFG populations and demonstrate that such emission can be characterized by just a small number of physically-motivated parameters. We further show that source populations would leave anisotropic signatures in the EGB, which could be used to yield information about the underlying properties, dynamics and evolution of CR-rich SFGs.}

\FullConference{37$^{\rm{th}}$ International Cosmic Ray Conference (ICRC 2021)\\
		July 12th -- 23rd, 2021\\
		Online -- Berlin, Germany}

\begin{document}
\maketitle

\section{Introduction}

\noindent
Several nearby star-forming galaxies (SFGs) have been resolved in $\gamma$-rays~\cite{Ajello2020ApJ_SFG}, establishing such systems as a candidate source class for the extra-galactic $\gamma$-ray background (EGB). 
Their $\gamma$-ray emission is driven by an abundant reservoir of cosmic rays (CRs). These interact with interstellar gases to form pions ($\pi^0$, $\pi^{\pm}$), with $\gamma$-ray photons being released by $\pi^0$ decays.
The CRs within these systems are presumably accelerated by diffusive shock acceleration processes, boosting low energy charged seed particles to relativistic energies, which would be operating in the abundance of violent shocked astrophysical environments that would emerge soon after the onset of star-formation. 

The EGB can be decomposed into two components: (1) the $\gamma$-ray emission arising from \textit{resolved} extra-galactic source populations, and (2) an isotropic component comprised of the accumulated emission from all \textit{unresolved} $\gamma$-ray emitting sources beyond our Galaxy, extending as far as the observable Universe. 
The contribution to the EGB from source classes such as blazars and AGN has been well studied~\citep[e.g.][]{Lamastra2017A&A}, with many such sources having been resolved (either directly using $\gamma$-rays, or by association with observations at other wavelengths). However, the exact balance of sources forming the EGB has not yet been firmly established. While it has been argued that the majority of the EGB flux originates in unresolved AGN sources, including blazars~\cite[e.g.][]{Ajello2015ApJ} and radio galaxies~\cite[e.g.][]{Stecker2019arXiv}, SFGs could still account for several tens of percent~\citep{Peretti2020MNRAS}.

Recently, the SFG contribution to the EGB was modeled using a prototype approach~\citep[see][]{Peretti2020MNRAS}, with $\gamma$-ray emission from source galaxies being weighted by their star-formation rates compared to the nearby starburst M82. Other studies developed physically-determined SFG template models~\cite[e.g.][which returned broadly similar EGB predictions as the M82 prototype method]{Owen2021MNRAS}. It was further considered~\cite[in][]{Owen2021MNRAS} that spatial anisotropies would be imprinted into the $\gamma$-ray sky
by a SFG population, and these were demonstrated to be sensitive to certain inherent properties of an source population of SFGs - in particular, the internal CR spectral index and the redshift distribution of the population. This work extends these earlier results, and further assesses the impact source galaxy properties would have in modifying the SFG contribution to the EGB and its imprinted spatial signatures.

\section{Emission of $\gamma$-rays from SFGs}
\label{sec:section_2}

\noindent
In SFGs, energetic hadronic CRs would predominantly interact through proton-proton (pp) pion-production processes~\citep{Owen2018MNRAS}.
This arises above a threshold proton kinetic energy of $T_{\rm p}^{\rm th} = 0.28~\text{GeV}$, with around 30\% of pions produced being $\pi^{0}$, from which
  $\gamma$-ray production proceeds (with a branching ratio of 98.8\% and timescale of $\sim 10^{-16}$ s) as
  $\pi^0 \rightarrow 2\gamma$. The weak energy-dependence of the pp interaction cross-section and the $\pi^0$ formation multiplicity yields a $\gamma$-ray spectrum closely tracing the shape of the underlying CR proton spectrum driving the emission. 

\subsection{CR spectrum and energy budget}
\label{sec:section_2_1}

\noindent
The internal CR spectrum of a SFG is typically well-described by a simple power-law. The spectral index $\Gamma$ has been found range from -1.9 to -2.3 in nearby SFGs detected in $\gamma$-rays~\citep[e.g.][]{Ajello2020ApJ_SFG}. Here, we adopt a proton spectrum of $\Gamma = -2.1$, i.e. the mid-value of this range. The CR proton density within a SFG is estimated
following the approach of~\cite{Owen2021MNRAS}. This can be parameterised using just five quantities: the star-formation rate of the host galaxy, $\mathcal{R}_{\rm SF}$, the size of the nuclear starburst region, $R$, the CR spectral index, $\Gamma$, the maximum CR proton energy, $E_{\rm max}$ and an advective escape fraction $f_{\rm adv}$, which is the fraction of CRs that would be removed from a SFG by advection in galactic outflows. Large-scale galactic outflows are common in distant, young SFGs~\cite[see, e.g.][]{Frye2002ApJ} and are driven by the confluence of feedback from the concentrated starburst episode arising in galactic cores. The impact of these outflows on the steady-state CR energy density in a galaxy can be substantial, reducing it by between a few and a few tens of percent~\cite{Peretti2019MNRAS}. We discuss our model for $f_{\rm adv}$ in section~\ref{sec:f_adv}. In the following calculations, we fix $R=0.1$ and $E_{\rm max} = 50\;\!{\rm PeV}$~\cite{Owen2021MNRAS}. $\mathcal{R}_{\rm SF}$ is specified by the adopted star-formation rate function (SFRF) for the population~\cite{Katsianis2017MNRAS}. 

\subsection{Template emission model}
\label{sec:section_2_2}
   
\noindent   
The volumetric rate at which pp interactions arise is given by $\dot{n}_{\rm p\pi}(\gamma_{\rm p}) = \langle n_{\rm H}\rangle \;\! n_{\rm p}(\gamma_{\rm p}) \;\!{c}\;\!\sigma_{\rm p\pi}(\gamma_{\rm p})$
\citep{Owen2021MNRAS}, where $\langle n_{\rm H}\rangle$ is the average ambient gas density within a SFG, $n_{\rm p}$ is the CR proton density (which we set as 1 cm$^{-3}$), and $\sigma_{\rm p\pi}$ is the total inelastic pp interaction cross-section. This is well-parametrised by
   \citep{Kafexhiu2014}. The production rate of $\gamma$-rays, which relies on the formation and subsequent decay of neutral pions, is described by the differential $\gamma$-ray inclusive cross section of the ${\rm pp} \rightarrow {\rm pp} \pi^0$ interaction channel. This is also well-parametrised to an accuracy of better than 10 per cent by~\cite{Kafexhiu2014}.
The emitted $\gamma$-rays can be attenuated by 
$\gamma\gamma$ pair-production between high-energy $\gamma$-rays and a low energy target photons provided by the CMB, starlight or dust-reprocessed starlight. This proceeds as $\gamma + \gamma \rightarrow e^+ + e^-$
at a rate $\dot{N}_{\gamma\gamma}$ 
determined by the spectral number density of target photons and the effective 
$\gamma\gamma$ interaction cross-section~\citep[see, e.g.][]{Gould1967PhRv}.
We define the characteristic $\gamma$-ray path length in a radiation field as $\ell_{\gamma\gamma}(\epsilon_{\rm \gamma}, x) = c/\dot{N}_{\gamma\gamma}(\epsilon_{\rm \gamma}, x)$. This is the distance over which an interaction would typically arise under conditions specified at location $x$ \citep[see][]{Gould1967PhRv}, which 
can be used to quantify the $\gamma$-ray attenuation factor within a galaxy. When averaged through an extended SFG source (modeled as a uniformly attenuating sphere of radius $R$), this may be approximated by a characteristic attenuation factor specified by the size of the star-forming nucleus $R$ and the (energy-dependent) effective $\gamma$-ray path length as 
$\mathcal{A}(\zeta) = \exp\left(-{\zeta^2}\right)$~\cite{Owen2021MNRAS}, where $\zeta = (R/\ell_{\gamma\gamma})^{1/2}$.

\begin{figure}
    \centering
    \includegraphics[width=0.95\columnwidth]{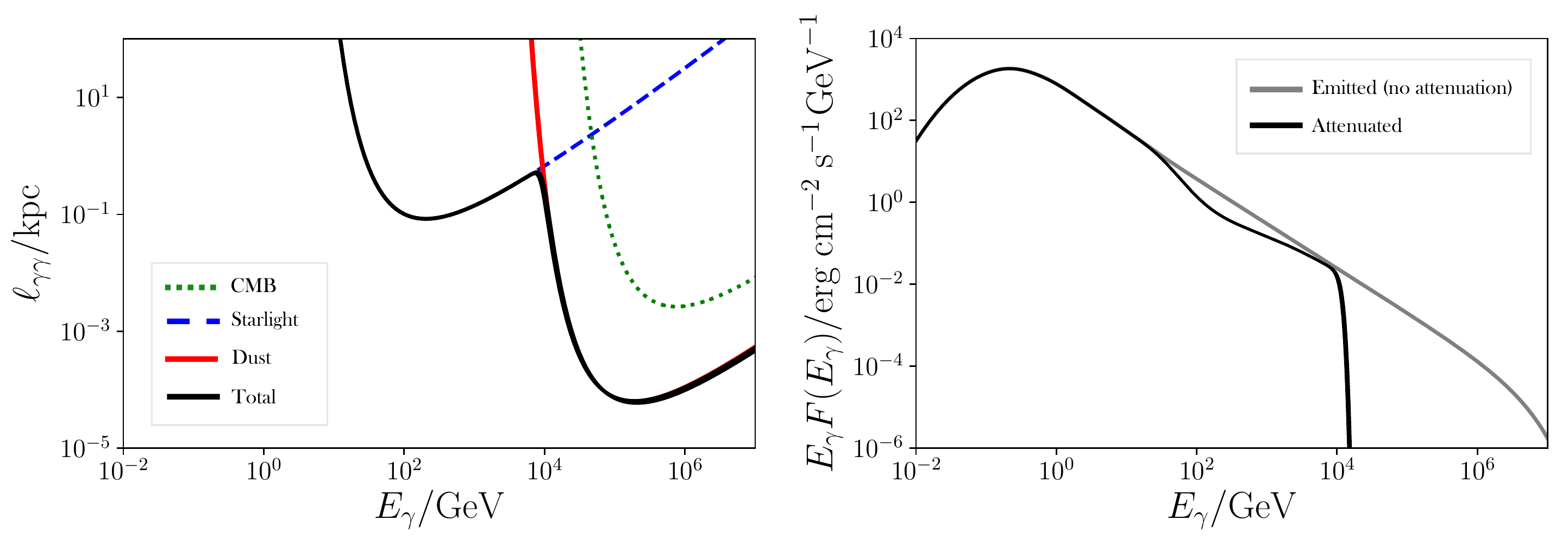}   
        \caption{Adapted from~\cite{Owen2021MNRAS}, with $\mathcal{R}_{\rm SF} = 10~\text{M}_{\odot}\;\!\text{yr}^{-1}$, $R = 0.1~\text{kpc}$ and at $z=2$. \textbf{Left}: Effective path lengths of $\gamma$-rays within a SFG model. Attenuation is strongly dominated by dust emission from $\sim$ 10 TeV.  
    {\bf Right}: $\gamma$-ray emission from the pp interaction is shown in grey. The black line shows the effective $\gamma$-ray emission spectrum emitted from the model SFG, accounting for $\gamma$-ray absorption within the SFG nucleus.}
  \label{fig:path_lengths}
\end{figure}

We consider that $\gamma$-ray attenuation within a SFG is dominated by three radiation fields: (1) the cosmological microwave background (CMB); (2) the interstellar radiation field (ISRF) from stars, and (3) the re-processed ISRF by interstellar dust. These may each be approximated with a blackbody spectrum defined by their characteristic temperature. We follow the treatment in~\cite{Owen2021MNRAS}, and the attenuating effects of these radiation fields on the emitted spectrum are demonstrated in Figure~\ref{fig:path_lengths} for a template galaxy example with $\mathcal{R}_{\rm SF} = 10~\text{M}_{\odot}\;\!\text{yr}^{-1}$, $R = 0.1~\text{kpc}$ and located at a redshift of $z=2$.

\subsection{Galaxy clustering}
\label{sec:section_2_3}

\noindent
The spatial distribution of SFGs, described by their power spectrum $P_{\rm g}(k, z)$, is reflected in their emission imprinted into the EGB. The bias of galaxy population clustering compared to that of dark matter is typically studied observationally from their spatial distribution, with various source classes having been found to exhibit different clustering properties~\cite[e.g. see][which finds a different clustering bias for AGNs and SFGs against dark matter]{Hale2018MNRAS}. 
To model this bias, we assume linear clustering of SFGs, and adopt a galaxy clustering length of $r_0^{\rm c} = 6.1\;\!\text{Mpc}\;\!h^{-1}$, and 
a power-law slope of the two-point correlation function of galaxies of $\iota = 1.8$~\cite{Hale2018MNRAS}. 

\subsection{Outflows and CR propagation in SFGs}
\label{sec:f_adv}

\noindent
The impact of outflows, which are ubiquitous among SFGs, in removing CRs from a galaxy has been argued to be substantial~\cite{Peretti2019MNRAS}, and this could have significant impacts for the resulting $\gamma$-ray emission from a source population. \cite{Owen2021MNRAS} accounted for this by simply introducing a CR reduction factor of 50\% uniformly across all SFGs. 
However, recent works have demonstrated that outflow velocities can evolve over redshift, particularly up to $z~\sim 2$~\cite{Sugahara2017ApJ}, with additional dependencies on $\mathcal{R}_{\rm SF}$ and the stellar mass of a host galaxy. This would modify the CR containment fraction compared to that assumed in the earlier model of~\cite{Owen2021MNRAS}. We introduce a simple CR containment model, where the CR spectral normalization is set by the advective loss fraction:\footnote{This approximation would not be valid at very high energies, for which faster CR diffusion would not effectively `contain' CRs in a host galaxy/outflow system.} 
\begin{equation}
    f_{\rm adv}(E_{\rm p}) \approx \left\{\frac{V_{\rm f}(\ell)\;\!\tau_{\rm adv}^{-1}(\ell)}{\tau_{\rm d}^{-1}(E_{\rm p})+\tau_{\rm adv}^{-1}(\ell)}\right\}\Bigg\vert_{\ell = \ell_0} \ .
\end{equation}
Here, $V_{\rm f}(\ell)$ is the volume filling factor of an outflow as a function of height $\ell$ from the starburst nucleus, $\tau_{\rm d}(E_{\rm p}) = \ell^2/4 D(E_{\rm p})$ is the characteristic CR diffusion timescale, and $\tau_{\rm adv} = \ell/v(\ell) \sim \ell/v_{\infty}$ is the characteristic CR advection timescale in an outflow. $v(\ell_0)\approx v_{\infty}$ is the velocity of the outflow at some point that may be considered as the interstellar boundary, beyond which the flow is approximated to have reached its terminal velocity. We set this point as $\ell_0 = 1\;\!{\rm kpc}$.

Some studies have considered that $V_{\rm f}$ may depend on redshift~\citep{Bertone2005MNRAS}, however the exact dependency remains unclear and presumably would reflect the detailed physical configuration of an outflow and its immediate environment. As such, we follow~\cite{KadoFong2020ApJ}, and adopt a value of 0.6, corresponding to the volume filling factor of the hot wind fluid at around $\ell_0 = 1\;\!{\rm kpc}$. 
We only consider variation of the diffusion coefficient through its energy-dependence, which takes the parametric form
$D(\gamma_{\rm p}) = D_0 \left( {r_L(\gamma_{\rm p},\langle |B| \rangle)|}/{r_{L,0}}\right)^{\varsigma}$,
where $\langle |B| \rangle = |B|$ is the characteristic local magnetic field strength. The normalization value $D_0 = 3.0\times 10^{28}$ cm$^2$ s$^{-1}$ is based on empirical measurements of the diffusion of CRs in the Milky Way, and is appropriate for a 1 GeV CR proton diffusing through a 5$\mu$G ISM magnetic field with gyro-radius $r_{L, 0}$. 

If assuming the same volume filling factor $V_{\rm f} = 0.6$, and that all SFGs are affected equally by outflows, the baseline model of~\cite{Owen2021MNRAS} would correspond to an outflow velocity of $\sim 80\;\!{\rm km}\;\!{\rm s}^{-1}$, i.e. significantly less than observationally determined ranges of values~\cite[e.g.][]{Sugahara2017ApJ}. Indeed, this choice was intended to indicate an upper estimate for the EGB intensity at $z=0$ but, if taken strictly, it would imply either substantially smaller filling factors were implicitly assumed, or only a subset of SFGs would host sufficiently strong outflows to advect CRs. In this work, we consider the impact a more physically-informed treatment of CR transport would have on the production of $\gamma$-rays in SFGs, accounting for the evolutionary trends of galactic outflow velocities over redshift described by~\cite{Sugahara2017ApJ}, using the scaling relation
\begin{equation}
    v_{\infty}(z, M^{\star}) \approx 320 \;\! \left(\frac{{\rm sSFR}}{0.316\; {\rm Gyr}^{-1}}\right)^{\beta_{\rm v}(z)}\;\!{\rm km}\;{\rm s}^{-1} \ ,
\end{equation}
where ${\rm sSFR} = \mathcal{R}_{\rm SF}/M_{\star}$ is the specific star-formation rate of a galaxy, and $\mathcal{R}_{\rm SF}$ is set by the redshift and stellar mass of a galaxy. 
We consider three forms for the function $\beta_{\rm v}(z)$, inspired by the best-fit values of~\cite{Sugahara2017ApJ}, where \textbf{Model 1} reflects their full sample of $z<0.2$ to $z=2.2$ galaxies with $\beta_{\rm v}(z) = 0.46$; \textbf{Model 2} reflects their restricted sample of galaxies up to $z=1.4$, where $\beta_{\rm v}(z) = 0.58$; and \textbf{Model 3} reflects the minimal redshift evolution scenario above $z=2$~\cite{Sugahara2019ApJ} such that $\beta_{\rm v}(z) = 0.58 \; \mathcal{H}(2-z)$.\footnote{$\mathcal{H}$ is the Heaviside step function.}

\section{EGB model}
\label{sec:section_3}

\subsection{Cosmological propagation of $\gamma$-rays}

\noindent
Over cosmological distances, the radiative transfer equation for $\gamma$-rays, in terms of redshift, is:
\begin{equation}
\frac{{\rm d}\mathcal{I}_{\gamma}}{{\rm d}z} = (1+z) \left[-\alpha_{\gamma \gamma} \mathcal{I}_{\gamma} + \frac{j_{\gamma}}{\nu^3} \right] \frac{{\rm d}s}{{\rm d}z}
\label{eq:radiative_transfer}
\end{equation}
where all quantities are Lorentz invariant, i.e. $\mathcal{I}_{\gamma} = I_{\gamma}/v^3$ for $I_{\gamma}$ as the local `proper' intensity (such that, in practice, co-moving absorption $\alpha_{\gamma \gamma}$ and emission $j_{\gamma}$ functions are used for the attenuation and emission of $\gamma$-rays respectively, and $\nu$ is co-moving frequency). ${{\rm d}s}/{{\rm d}z}$ is defined for a flat Friedmann-Robertson-Walker (FRW) Universe. 
We solve Equation~\ref{eq:radiative_transfer} numerically over redshift to model the SFG contribution to the EGB spectrum at $z=0$, with the SFG source population (using~\cite{Katsianis2017MNRAS}) distributed up to $z_{\rm max} = 3$.\footnote{This range covers the peak of cosmic star-formation, and would presumably account for the majority of $\gamma$-ray emission from SFGs.}

Over cosmological distances, $\gamma$-rays are reprocessed by pair-production in the extra-galactic background light (EBL) and subsequent inverse-Compton scattering off CMB photons (the $\gamma$-ray \textit{cascade}). The attenuating part of this process can be characterized by
an optical depth: 
\begin{equation}
\tau_{\gamma \gamma}(z, \epsilon_{\gamma}) \equiv \;\!\int_0^z \alpha_{\gamma \gamma}(z', \epsilon_{\gamma}) \;\! \frac{{\rm d}s}{{\rm d}z'}{\rm d}z'  \ .
\end{equation}
The associated cascade emission is computed assuming inverse-Compton scattering of pair-produced electrons dominates, and this is encoded by the term $j_{\gamma}$ together with direct emission from SFGs. 
If adopting the the semi-analytic model of~\cite{Inoue2013ApJ}, for which $\gamma$-ray optical depths for the EBL are provided between 1 GeV and 45 TeV, and up to a redshift of $z=10$, the $\gamma$-ray absorption coefficient $\alpha_{\gamma\gamma}$ can be computed from the differential optical depth as a function of redshift.

\begin{figure}
    \centering
    \includegraphics[width=0.8\columnwidth]{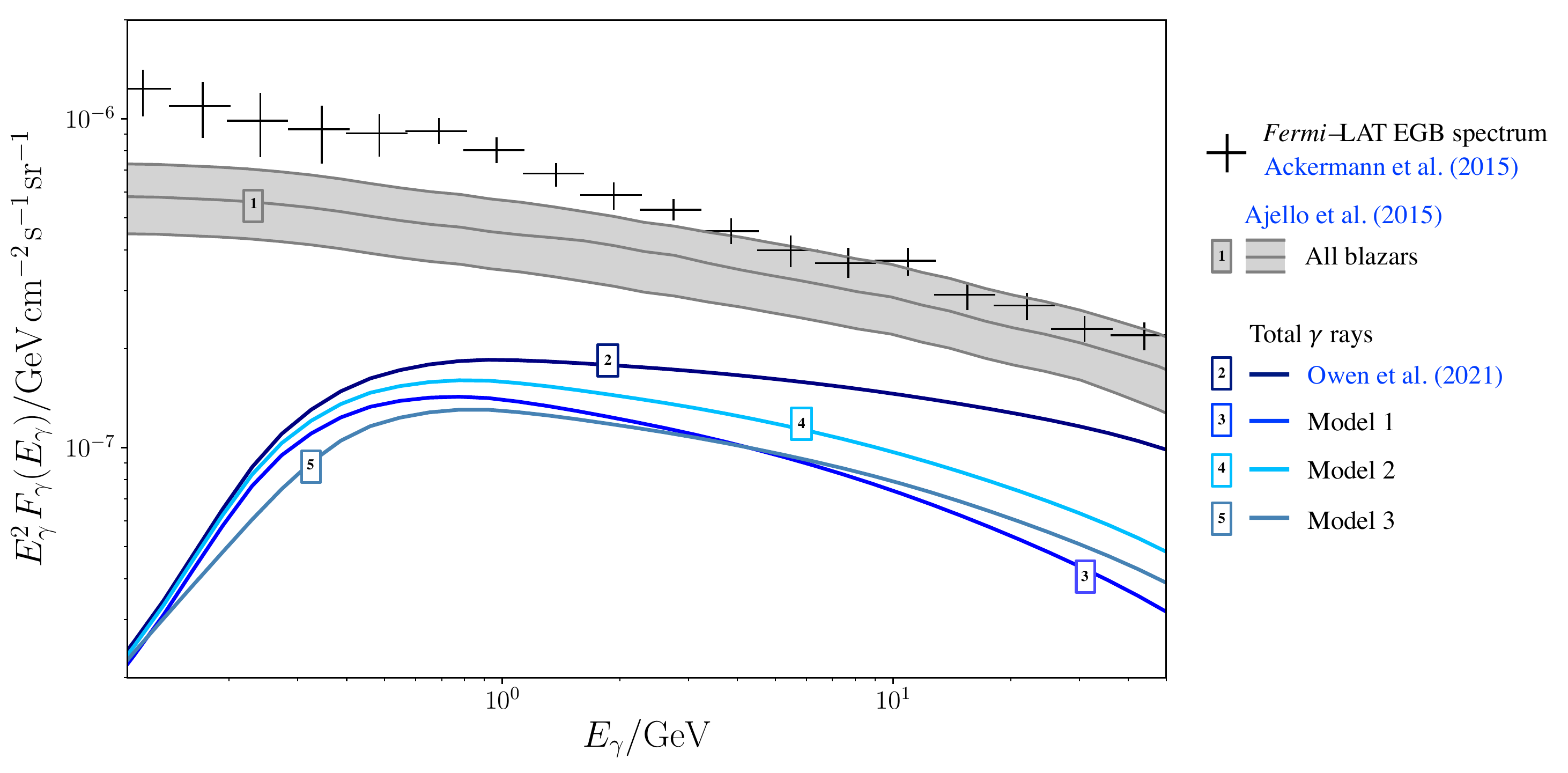}   
        \caption{EGB spectrum from SFGs under the 3 outflow redshift-dependency models. While CR advection in outflows is important at lower energies, diffusive propagation starts to become more competitive at higher energies. For all 3 outflow models, the EGB flux is reduced at all energies compared to the results of \textcolor{blue}{Owen et al. 2021}~\cite{Owen2021MNRAS}, however this reduction is more substantial at higher energies. The band in grey, labeled as line 1, represents the `all blazar' contribution of \textcolor{blue}{Ajello et al. 2015}~\cite{Ajello2015ApJ}, while the \textit{Fermi}-LAT data points are from \textcolor{blue}{Ackermann et al. 2015}~\citep{Ackermann2015ApJ}.}
  \label{fig:final_spec}
\end{figure}

We solve equation~\ref{eq:radiative_transfer} for a fixed outflow loss fraction of 50\% (thus reproducing the result of~\cite{Owen2021MNRAS}), and for the three outflow velocity models discussed in section~\ref{sec:f_adv}. The resulting EGB spectrum is shown in  Figure~\ref{fig:final_spec}, which shows the choice of CR transport model to be important -- particularly at higher energies, where diffusive CR propagation becomes more significant.

\subsection{EGB anisotropy}

\noindent
The SFG power spectrum $P_{\rm g}(k, z)$ (see section~\ref{sec:section_2_3}) would imprint a spatial signature in the EGB, even though individual contributing sources would not typically be resolved. The distribution of spatial scales of this signature would depend on redshift $z$, being specified by $P_{\rm g}(k, z)$ and the strength of the contribution from sources at a particular epoch. 
The imprinted signatures could be accessed in the EGB using the intensity fluctuation power spectrum $\mathcal{C}_{\ell}$, providing an alternative window to study EGB source populations and their variation over $z$. 

\begin{figure}
\floatbox[{\capbeside\thisfloatsetup{capbesideposition={right,center},capbesidewidth=6cm}}]{figure}[\FBwidth]
{\caption{EGB intensity fluctuation power spectrum $\mathcal{C}_{\ell}$, normalized to $\mathcal{C}_{10}$, plotted against multipole $\ell$ in the energy band $E_{\gamma} = (1-10) \;\!{\rm GeV}$. This shows the result of \textcolor{blue}{Owen et al. 2021} \cite{Owen2021MNRAS}, compared to the 3 outflow models, as labeled, which all show a slightly broader power spectrum. This is most noticeable in the case of Model 3, where the broadening is stronger, particularly towards larger scales (smaller $\ell$).}
\label{fig:anisotropy}}
{\includegraphics[width=7cm]{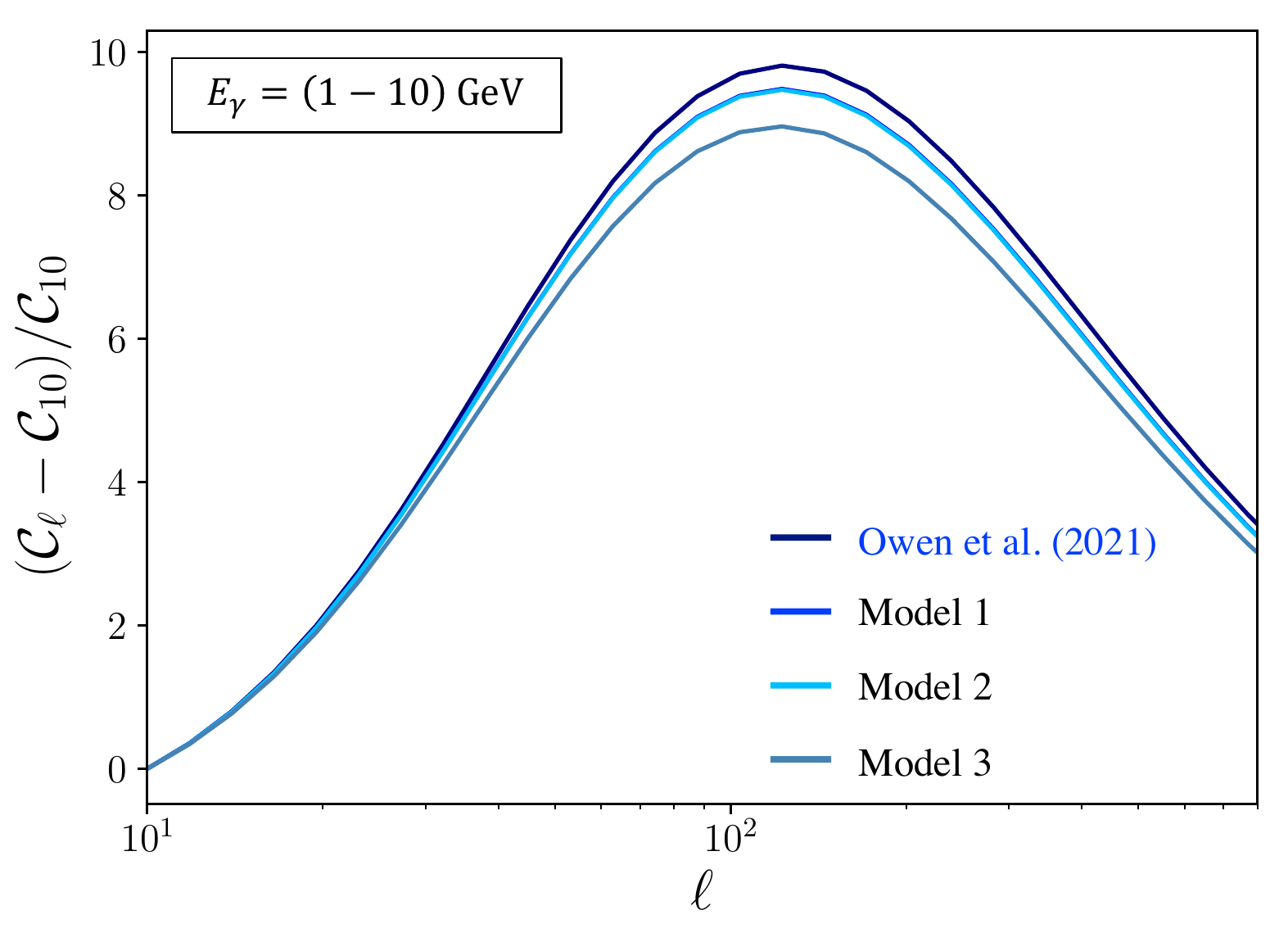}}
\end{figure}

With sufficiently detailed modeling of signatures, crucial insights about the CR activity in EGB source populations could be uncovered from anisotropies in the $\gamma$-ray sky. As a demonstration, Figure~\ref{fig:anisotropy} shows how the EGB anisotropy signature differs between the four source models considered in this work. Notably, Model 3 (which introduces an additional redshift dependence compared to the other models) shows a different shape of EGB anisotropy signature, indicating how underlying differences in source populations and their evolution may be probed. Such signatures would be accessible in the near-future with \textit{Fermi}-LAT and the up-coming Cherenkov Telescope Array~\cite{Owen2021MNRAS}.
 
\section{Summary and conclusions}
\label{sec:section_4}

\noindent
This work has shown how signatures are imprinted into the EGB by SFG populations, and that these can be modeled using a template approach, specified by just a small number of physically-motivated parameters. The resulting EGB spectrum was shown to be sensitive to source population properties, in particular CR transport processes in SFGs which must be carefully modeled in future work to robustly determine their exact EGB contributions. Information about the redshift distribution of EGB sources was shown to be accessible in the EGB intensity fluctuation power spectrum which, with detailed modeling, could be used to uncover the redshift distribution of contributing source populations and even reveal details about their evolutionary characteristics. This could have important applications in resolving CR activity in galaxies during the `high noon' of cosmic star-formation. 

\acknowledgments

\noindent
This work used high-performance computing facilities operated by the
Center for Informatics and Computation in Astronomy (CICA) at National
Tsing Hua University (NTHU). This equipment was funded by the Ministry of
Education of Taiwan and the Ministry of Science and Technology of Taiwan (MOST).
ERO is supported by the Ministry of Education of Taiwan at CICA, NTHU. KGL acknowledges support from JSPS KAKENHI grants JP18H05868 and JP19K14755. AKHK acknowledges support from MOST (grant 109-2628-M-007-005-RSP).

\bibliographystyle{ICRC}
\bibliography{references}

\providecommand{\href}[2]{#2}\begingroup\raggedright\begin{thebibliography}{10}

\bibitem{Ajello2020ApJ_SFG}
M.~{Ajello}, M.~{Di Mauro} et~al., \emph{{The {\ensuremath{\gamma}}-Ray
  Emission of Star-forming Galaxies}},
  \href{https://doi.org/10.3847/1538-4357/ab86a6}{\emph{ApJ} {\bfseries 894}
  (2020) 88} [\href{https://arxiv.org/abs/2003.05493}{{\ttfamily 2003.05493}}].

\bibitem{Lamastra2017A&A}
A.~{Lamastra}, N.~{Menci} et~al., \emph{{Extragalactic gamma-ray background
  from AGN winds and star-forming galaxies in cosmological galaxy-formation
  models}}, \href{https://doi.org/10.1051/0004-6361/201731452}{\emph{A\&A}
  {\bfseries 607} (2017) A18}
  [\href{https://arxiv.org/abs/1709.03497}{{\ttfamily 1709.03497}}].

\bibitem{Ajello2015ApJ}
M.~{Ajello}, D.~{Gasparrini} et~al., \emph{{The Origin of the Extragalactic
  Gamma-Ray Background and Implications for Dark Matter Annihilation}},
  \href{https://doi.org/10.1088/2041-8205/800/2/L27}{\emph{ApJ} {\bfseries 800}
  (2015) L27} [\href{https://arxiv.org/abs/1501.05301}{{\ttfamily
  1501.05301}}].

\bibitem{Stecker2019arXiv}
F.W.~{Stecker}, C.R.~{Shrader} et~al., \emph{{The Extragalactic Gamma-Ray
  Background from Core-dominated Radio Galaxies}},
  \href{https://doi.org/10.3847/1538-4357/ab23ee}{\emph{ApJ} {\bfseries 879}
  (2019) 68} [\href{https://arxiv.org/abs/1903.06544}{{\ttfamily 1903.06544}}].

\bibitem{Peretti2020MNRAS}
E.~{Peretti}, P.~{Blasi} et~al., \emph{{Contribution of starburst nuclei to the
  diffuse gamma-ray and neutrino flux}},
  \href{https://doi.org/10.1093/mnras/staa698}{\emph{MNRAS} {\bfseries 493}
  (2020) 5880} [\href{https://arxiv.org/abs/1911.06163}{{\ttfamily
  1911.06163}}].

\bibitem{Owen2021MNRAS}
E.R.~{Owen}, K.-G.~{Lee} et~al., \emph{{Characterizing the signatures of
  star-forming galaxies in the extragalactic {\ensuremath{\gamma}}-ray
  background}}, \href{https://doi.org/10.1093/mnras/stab1707}{\emph{MNRAS}
  {\bfseries 506} (2021) 52}
  [\href{https://arxiv.org/abs/2106.07308}{{\ttfamily 2106.07308}}].

\bibitem{Owen2018MNRAS}
E.R.~{Owen}, I.B.~{Jacobsen} et~al., \emph{{Interactions between
  ultra-high-energy particles and protogalactic environments}},
  \href{https://doi.org/10.1093/mnras/sty2279}{\emph{MNRAS} {\bfseries 481}
  (2018) 666} [\href{https://arxiv.org/abs/1808.07837}{{\ttfamily
  1808.07837}}].

\bibitem{Frye2002ApJ}
B.~{Frye}, T.~{Broadhurst} et~al., \emph{{Spectral Evidence for Widespread
  Galaxy Outflows at z > 4}}, \href{https://doi.org/10.1086/338965}{\emph{ApJ}
  {\bfseries 568} (2002) 558}
  [\href{https://arxiv.org/abs/astro-ph/0112095}{{\ttfamily
  astro-ph/0112095}}].

\bibitem{Peretti2019MNRAS}
E.~{Peretti}, P.~{Blasi} et~al., \emph{{Cosmic ray transport and radiative
  processes in nuclei of starburst galaxies}},
  \href{https://doi.org/10.1093/mnras/stz1161}{\emph{MNRAS} {\bfseries 487}
  (2019) 168} [\href{https://arxiv.org/abs/1812.01996}{{\ttfamily
  1812.01996}}].

\bibitem{Katsianis2017MNRAS}
A.~{Katsianis}, G.~{Blanc} et~al., \emph{{The evolution of the star formation
  rate function in the EAGLE simulations: a comparison with UV, IR and
  H{\ensuremath{\alpha}} observations from z {\ensuremath{\sim}} 8 to z
  {\ensuremath{\sim}} 0}},
  \href{https://doi.org/10.1093/mnras/stx2020}{\emph{MNRAS} {\bfseries 472}
  (2017) 919} [\href{https://arxiv.org/abs/1708.01913}{{\ttfamily
  1708.01913}}].

\bibitem{Kafexhiu2014}
E.~{Kafexhiu}, F.~{Aharonian} et~al., \emph{{Parametrization of gamma-ray
  production cross sections for p p interactions in a broad proton energy range
  from the kinematic threshold to PeV energies}},
  \href{https://doi.org/10.1103/PhysRevD.90.123014}{\emph{PhRvD} {\bfseries 90}
  (2014) 123014} [\href{https://arxiv.org/abs/1406.7369}{{\ttfamily
  1406.7369}}].

\bibitem{Gould1967PhRv}
R.J.~{Gould} and G.P.~{Schr{\'e}der}, \emph{{Pair Production in Photon-Photon
  Collisions}}, \href{https://doi.org/10.1103/PhysRev.155.1404}{\emph{Physical
  Review} {\bfseries 155} (1967) 1404}.

\bibitem{Hale2018MNRAS}
C.L.~{Hale}, M.J.~{Jarvis} et~al., \emph{{The clustering and bias of
  radio-selected AGN and star-forming galaxies in the COSMOS field}},
  \href{https://doi.org/10.1093/mnras/stx2954}{\emph{MNRAS} {\bfseries 474}
  (2018) 4133} [\href{https://arxiv.org/abs/1711.05201}{{\ttfamily
  1711.05201}}].

\bibitem{Sugahara2017ApJ}
Y.~{Sugahara}, M.~{Ouchi} et~al., \emph{{Evolution of Galactic Outflows at
  z$\sim$ 0 -- 2 Revealed with SDSS, DEEP2, and Keck Spectra}},
  \href{https://doi.org/10.3847/1538-4357/aa956d}{\emph{ApJ} {\bfseries 850}
  (2017) 51} [\href{https://arxiv.org/abs/1703.01885}{{\ttfamily 1703.01885}}].

\bibitem{Bertone2005MNRAS}
S.~{Bertone}, F.~{Stoehr} et~al., \emph{{Semi-analytic simulations of galactic
  winds: volume filling factor, ejection of metals and parameter study}},
  \href{https://doi.org/10.1111/j.1365-2966.2005.08772.x}{\emph{MNRAS}
  {\bfseries 359} (2005) 1201}
  [\href{https://arxiv.org/abs/astro-ph/0402044}{{\ttfamily
  astro-ph/0402044}}].

\bibitem{KadoFong2020ApJ}
E.~{Kado-Fong}, J.-G.~{Kim} et~al., \emph{{Diffuse Ionized Gas in Simulations
  of Multiphase, Star-forming Galactic Disks}},
  \href{https://doi.org/10.3847/1538-4357/ab9abd}{\emph{ApJ} {\bfseries 897}
  (2020) 143} [\href{https://arxiv.org/abs/2006.06697}{{\ttfamily
  2006.06697}}].

\bibitem{Sugahara2019ApJ}
Y.~{Sugahara}, M.~{Ouchi} et~al., \emph{{Fast Outflows Identified in Early
  Star-forming Galaxies at z = 5-6}},
  \href{https://doi.org/10.3847/1538-4357/ab49fe}{\emph{ApJ} {\bfseries 886}
  (2019) 29} [\href{https://arxiv.org/abs/1904.03106}{{\ttfamily 1904.03106}}].

\bibitem{Inoue2013ApJ}
Y.~{Inoue}, S.~{Inoue} et~al., \emph{{Extragalactic Background Light from
  Hierarchical Galaxy Formation: Gamma-Ray Attenuation up to the Epoch of
  Cosmic Reionization and the First Stars}},
  \href{https://doi.org/10.1088/0004-637X/768/2/197}{\emph{ApJ} {\bfseries 768}
  (2013) 197} [\href{https://arxiv.org/abs/1212.1683}{{\ttfamily 1212.1683}}].

\bibitem{Ackermann2015ApJ}
M.~{Ackermann}, M.~{Ajello} et~al., \emph{{The Spectrum of Isotropic Diffuse
  Gamma-Ray Emission between 100 MeV and 820 GeV}},
  \href{https://doi.org/10.1088/0004-637X/799/1/86}{\emph{ApJ} {\bfseries 799}
  (2015) 86} [\href{https://arxiv.org/abs/1410.3696}{{\ttfamily 1410.3696}}].

\end{thebibliography}\endgroup

%
%
%

\end{document}